\documentclass{raa}        
\usepackage{graphicx,times}
\usepackage{natbib}
\usepackage{amssymb,amsmath}
\bibpunct{(}{)}{;}{a}{}{,}

\usepackage[a4paper=true,dvipdfm=true,pagebackref=true]{hyperref}
\hypersetup{pdftitle = The title of my PDF, pdfauthor = My name, pdfsubject= The subject, pdfkeywords = keyword1 keyword2 keyword3} 
\hypersetup{colorlinks = true, linkcolor = green, anchorcolor = red, citecolor = blue, filecolor = red, pagecolor = red, urlcolor = red}

\begin{document}

   \title{Improved proper motion determinations for 15 open clusters based on the UCAC4 catalog}

 \volnopage{ {\bf 2016} Vol.\ {\bf X} No. {\bf XX}, 000--000}
   \setcounter{page}{1}

   \author{Alexander Kurtenkov\inst{1,2}, Nadezhda Dimitrova\inst{3}, Alexander Atanasov\inst{4}, Teodor D. Aleksiev\inst{5}
   }

   \institute{ Department of Astronomy, University of Sofia, 5 James Bourchier Blvd., 1164 Sofia, 
Bulgaria; {\it al.kurtenkov@gmail.com}\\
        \and
             Institute of Astronomy and National Astronomical Observatory, Bulgarian Academy of Sciences, 72 Tsarigradsko Shose Blvd., 1784 Sofia, Bulgaria\\
	\and
             Anglo-American School of Sofia, 1 Siyanie Str., 1138 Sofia, Bulgaria\\
	\and
Geo Milev High School of Mathematics, Storgozia, 5802 Pleven, Bulgaria\\
\and 
Sofia High School of Mathematics, 61 Iskar Str., 1000 Sofia, Bulgaria\\
\vs \no
   {\small Received 2015 October 8; accepted 2016 March 2}
}

\abstract{The proper motions of 15 nearby ($d<1$ kpc) open clusters were recalculated using data from the UCAC4 catalog. Only evolved or main sequence stars inside a certain radius from the center of the cluster were used. The results differ significantly from the ones presented by Dias et al. (2014). This could be explained by the different approach to taking the field star contamination into account. The present work aims to emphasize the importance of applying photometric criteria for the calculation of OC proper motions.}

\keywords{proper motions --- Galaxy: open clusters and associations: general

}

   \authorrunning{A. Kurtenkov et al. }            
   \titlerunning{Improved proper motion determinations for 15 close open clusters based on the UCAC4 catalog}  
   \maketitle

%
\section{Introduction}           
\label{sect:intro}

Open clusters (OCs) are fundamental building blocks of spiral and irregular galaxies. Studies of galactic OCs have produced a vast amount of important scientific results in areas such as stellar evolution and star formation \citep{2002MNRAS.334..193C,1993AJ....106.1870P}. Furthermore, galactic OCs are crucial for the understanding of the structure and dynamics of the Milky Way. Open clusters and OB associations have been used to explore local structures \citep{1999AJ....117..354D,2000A&A...359...82T} as well as the large-scale structure of the galaxy \citep{2014MNRAS.437.1549B,2008ChJAA...8...96Z}. They also help tracing the chemical composition throughout the galactic thin disk \citep{2015arXiv150508027M}.

Proper motion is a key parameter of open clusters. Proper motions, distances and radial velocities are used to derive galactocentric velocities of OCs. The latter are of fundamental significance in galactic dynamics studies, e.g. determination of OC orbits \citep{2009MNRAS.399.2146W} and rotation of the Galaxy \citep{2005ApJ...629..825D,2007IAUS..235..150Z}. Another important implication of OC proper motions is the calculation of membership probabilities for individual stars \citep{1971A&A....14..226S,1985A&A...150..298C}. It has been shown that cluster parameters based on photometric membership probabilities are consistent with those based on proper motion membership probabilities, see e.g. \citet{2007AJ....133.2061W}.

The early history of open cluster proper motion determinations has been outlined by \citet{1962AJ.....67..699V}. Up until the end of the 20$^{th}$ century proper motions of OCs were derived mainly on a case-by-case basis. The first large catalog was compiled by \citet{1997AstL...23...71G}, for 181 clusters with $log(age)<8.3$. Large OC proper motion catalogs were later released by \citet{2000A&AS..146..251B} and \citet{2001A&A...376..441D,2002A&A...388..168D} using Hipparcos and Tycho-2 data respectively. The results, obtained by \citet{2003ARep...47....6L} were also based on the Tycho-2 catalog, and are currently the ones cited in the SIMBAD database. 

The UCAC4 catalog \citep{2013AJ....145...44Z} contains proper motion data for more than 105 million objects (complete to R=16\,mag). It compiles astrometric data from over 140 catalogs, including Hipparcos and Tycho-2, for the derivation of mean positions and proper motions. The astrometry is complemented by optical and NIR photometry from APASS and 2MASS. \citet{2014A&A...564A..79D} have used UCAC4 to obtain proper motions for 1805 galactic OCs. We have recalculated the proper motions of 15 close ($d<1\,kpc$ from the Sun) open clusters via a different method and obtained results, significantly different from the ones by \citet{2014A&A...564A..79D}.


\section{Object selection and method}

The open clusters for this work were selected from the WEBDA list\footnote{See \url{http://www.univie.ac.at/webda/dist_list.html}} of close OCs ($d<1\,kpc$). Clusters closer than $300\,pc$ were not included as there should be systematic differences between the proper motions of their members, depending on location. We chose only prominent OCs, whose color-magnitude diagrams (CMDs) present typical features for open clusters (main sequence, turnoff point). The selected clusters are presented in Table\,\ref{table_objects}.

\begin{table}
\begin{center}
\caption[Open clusters studied in the current work.]{Open clusters studied in the current work. The basic parameters are retrieved from the WEBDA database.}\label{table_objects}
\begin{tabular}{l@{}l@{}l@{ }c@{ }c@{ }c@{ } c@{} c@{} c@{} c@{}}
\hline\hline
cluster & alt. name \,& $\alpha(J2000)$  & $\delta(J2000)$  & l \, & b \, & dist. [pc] \,& (m-M) \,& E(B-V) \,& log(age)\, \\
\hline
 \object{NGC 1039} \,\, & M34 &  02:42:05    \, &    +42:45:42  \, &	 143.658 \, & -15.613 \, & 499  \, & 8.71  \, & 0.07 \, & 8.25  \\
 \object{NGC 1647} \,\, & $-$ & 04:45:55    \, &    +19:06:54  \, &	 180.337 \, & -16.772 \, & 540  \, & 9.81  \, & 0.37 \, & 8.16  \\
 \object{NGC 1662} \,\, & $-$ & 04:48:27    \, &    +10:56:12  \, &	 187.695 \, & -21.114 \, & 437  \, & 9.14  \, & 0.30 \, & 8.63  \\
 \object{NGC 2281} \,\, & $-$ & 06:48:17    \, &    +41:04:42  \, &	 174.901 \, & 16.881  \, & 558  \, & 8.93  \, & 0.06 \, & 8.55  \\
 \object{NGC 2358} \,\, & $-$ & 07:16:55    \, &    -17:08:59  \, &	 231.05  \, & -2.30   \, & 630  \, & 9.06  \, & 0.02 \, & 8.72  \\
 \object{NGC 2422} \,\, & M47 & 07:36:35    \, &    -14:29:00  \, &	 230.958 \, & 3.130   \, & 490  \, & 8.67  \, & 0.07 \, & 7.86  \\
 \object{NGC 2516} \,\, & $-$ & 07:58:04    \, &    -60:45:12  \, &	 273.816 \, & -15.856 \, & 409  \, & 8.37  \, & 0.10 \, & 8.05  \\
 \object{NGC 2547} \,\, & $-$ & 08:10:09    \, &    -49:12:54  \, &	 264.465 \, & -8.597  \, & 455  \, & 8.42  \, & 0.04 \, & 7.56  \\
 \object{NGC 3532} \,\, & $-$ & 11:05:39    \, &    -58:45:12  \, &	 289.571 \, & 1.347   \, & 486  \, & 8.55  \, & 0.04 \, & 8.49  \\
 \object{NGC 6124} \,\, & $-$ & 16:25:20    \, &    -40:39:12  \, &	 340.741 \, & 6.016   \, & 512  \, & 10.87 \, & 0.75 \, & 8.15  \\
 \object{NGC 6281} \,\, & $-$ & 17:04:41    \, &    -37:59:06  \, &	 347.731 \, & 1.972   \, & 479  \, & 8.86  \, & 0.15 \, & 8.50  \\
 \object{NGC 6405} \,\, & M6 & 17:40:20    \, &    -32:15:12  \, &	 356.580 \, & -0.777  \, & 487  \, & 8.88  \, & 0.14 \, & 7.97  \\
 \object{NGC 6494} \,\, & M23 & 17:57:04    \, &    -18:59:06  \, &	 9.894   \, & 2.834   \, & 628  \, & 10.09 \, & 0.36 \, & 8.48  \\
 \object{NGC 7092} \,\, & M39 & 21:31:48    \, &    +48:26:00  \, &	 92.403  \, & -2.242  \, & 326  \, & 7.61  \, & 0.01 \, & 8.45  \\
 \object{IC 4725}  \,\, & M25 & 18:31:47    \, &    -19:07:00  \, &    13.702  \, & -4.434  \, & 620  \, & 10.44 \, & 0.48 \, & 7.97  \\
\hline
\end{tabular} 
\end{center}  
\end{table}  

\begin{figure}
\centering
\includegraphics[width=14cm]{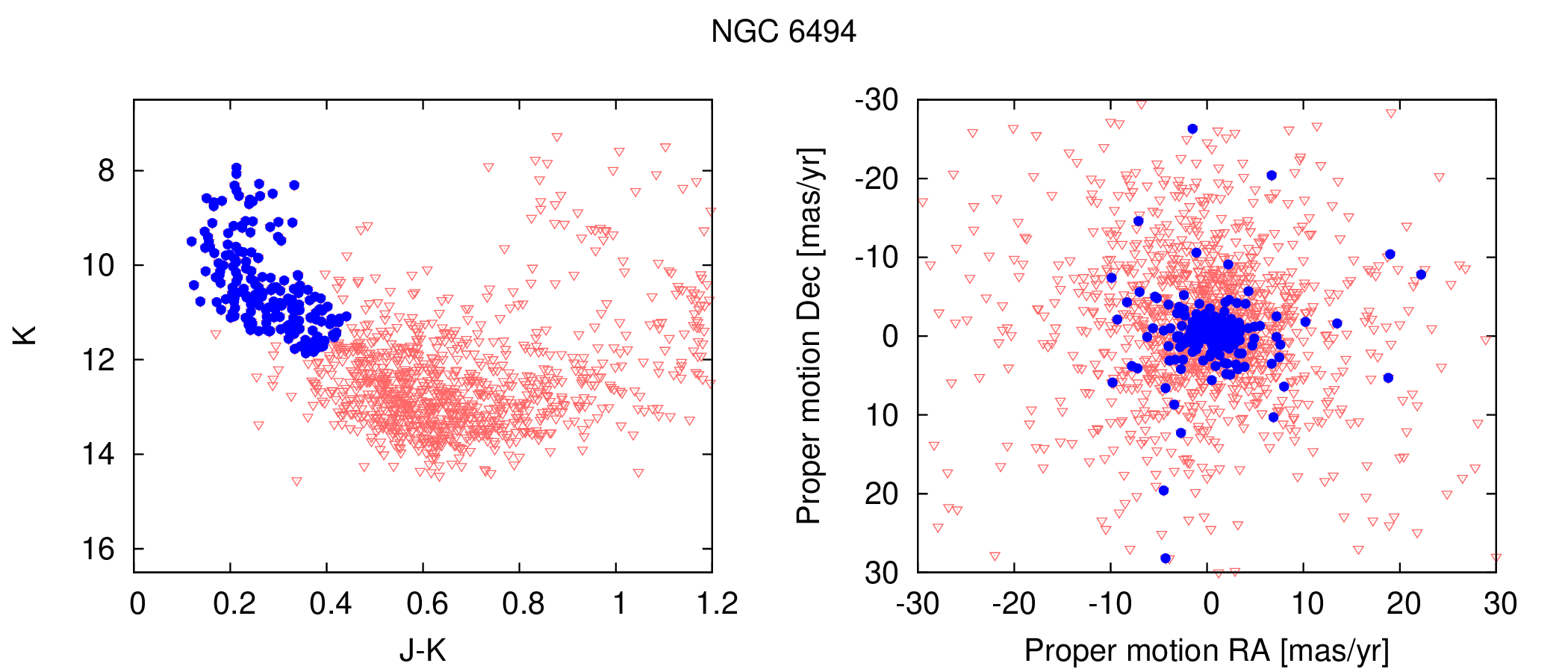}
\includegraphics[width=14cm]{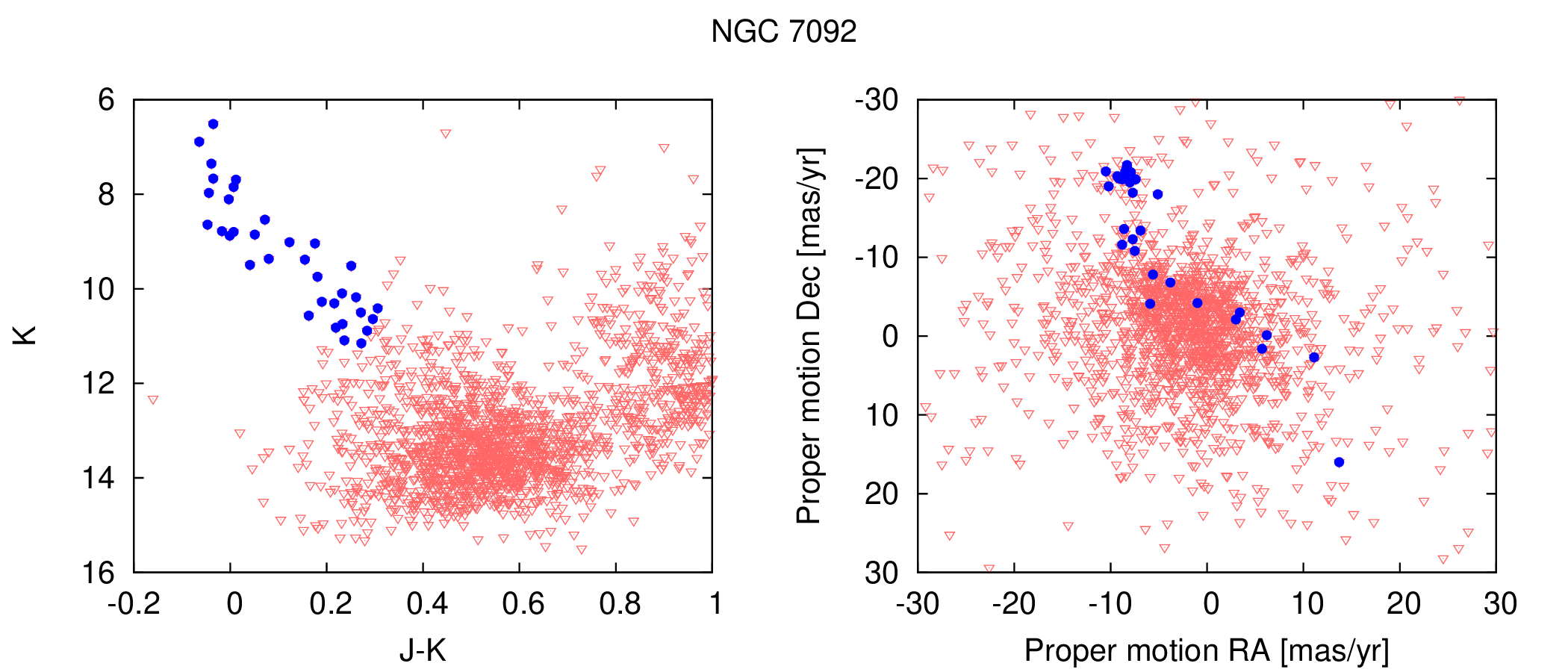}
\includegraphics[width=14cm]{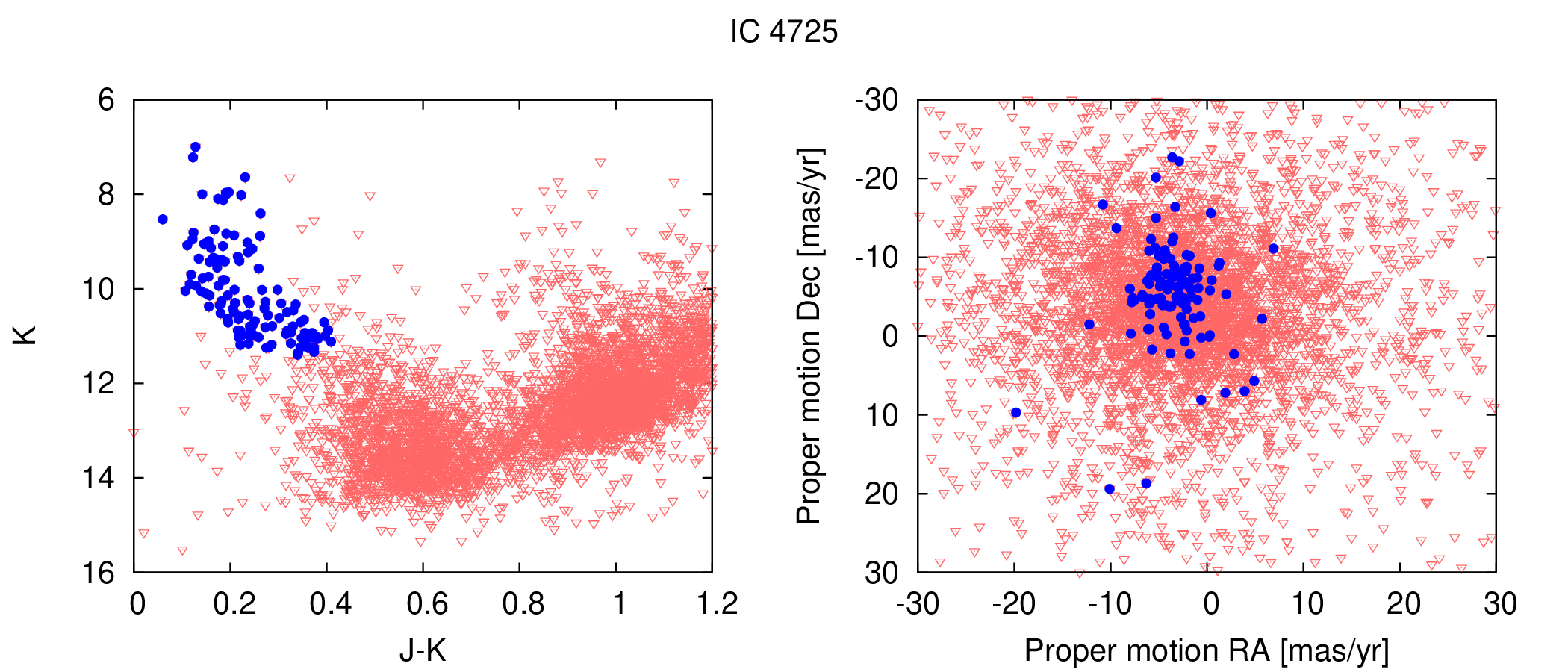}
\caption{\textit{left}: NIR color-magnitude diagrams of all clusters. All stars inside the search radii are plotted with red triangles. Filled blue dots represent highly probable cluster members (the $N_{1}$ subselections), used to calculate the OC proper motions after the ommission of outlying points. \textit{right}: All sources plotted in a $\mu_{\alpha}cos\delta$ vs $\mu_{\delta}$ plane. Cluster members appear grouped together.}
\label{CMDs}
\end{figure}

Stars in the vicinity of each cluster were extracted by searching the UCAC4 catalog inside a given radius from the cluster center. We used the same coordinates and radii of search as \citet{2014A&A...564A..79D}. A 2MASS $(J-K)$ vs $K$ diagram was built for each cluster. Out of all the $N_{0}$ stars, $N_{1}$ were selected as very probable cluster members based on their location on the CMD. Only stars lying on the main sequence (MS) or evolved ones, i.e. to the right from the MS and forming a feature along an isochrone, were included in the $N_{1}$ subselections (Fig.\ref{CMDs}). Data selection was carried out using Virtual Observatory tools (Aladin\footnote{See \url{http://aladin.u-strasbg.fr/}} and TOPCAT\footnote{See \url{http://www.star.bris.ac.uk/~mbt/topcat/}}).

\begin{figure}
\centering
\includegraphics[width=14cm]{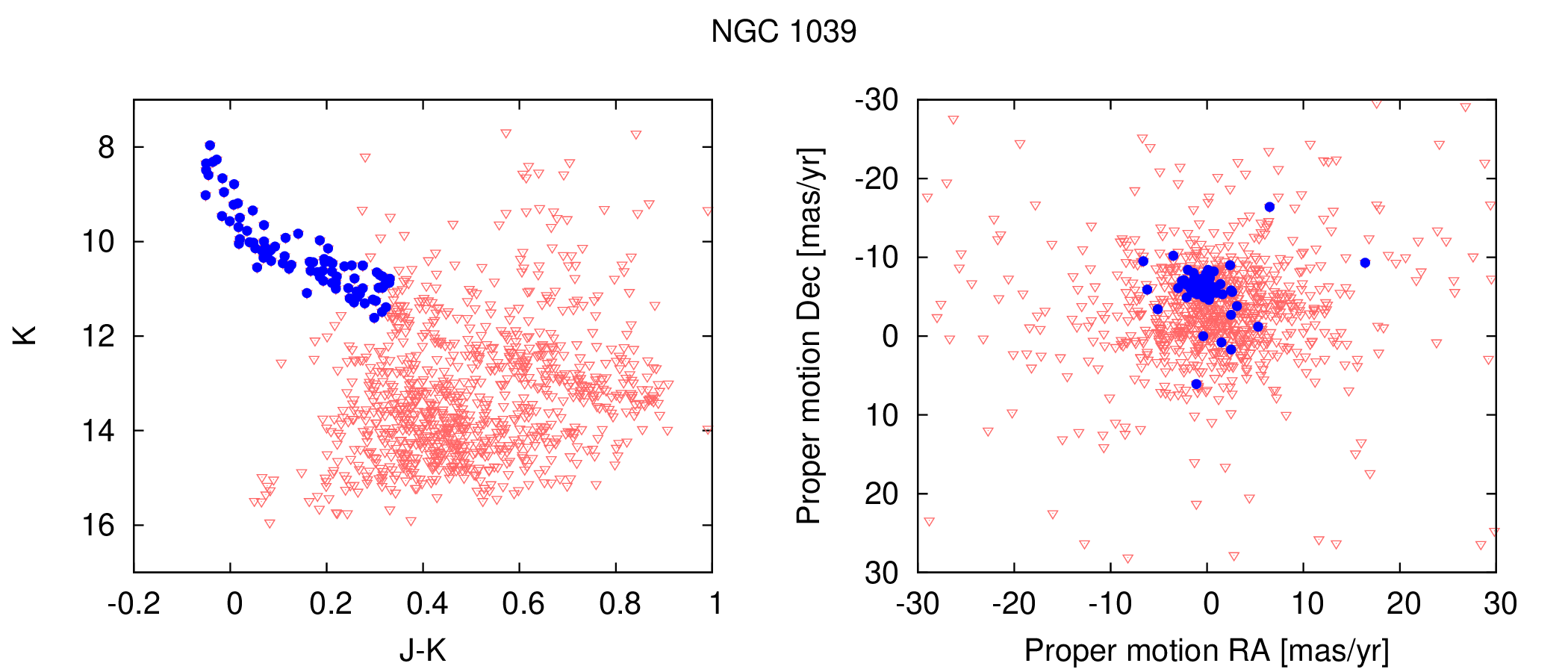}
\includegraphics[width=14cm]{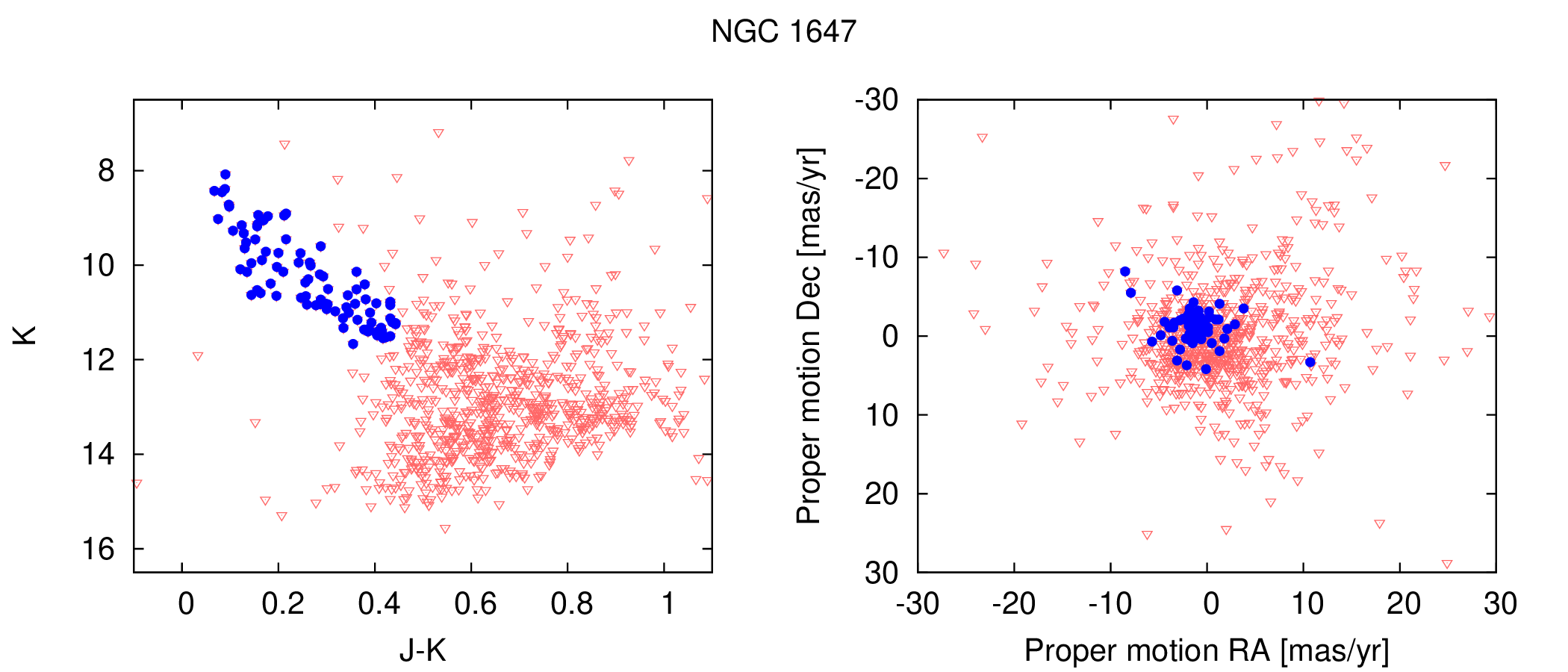}
\includegraphics[width=14cm]{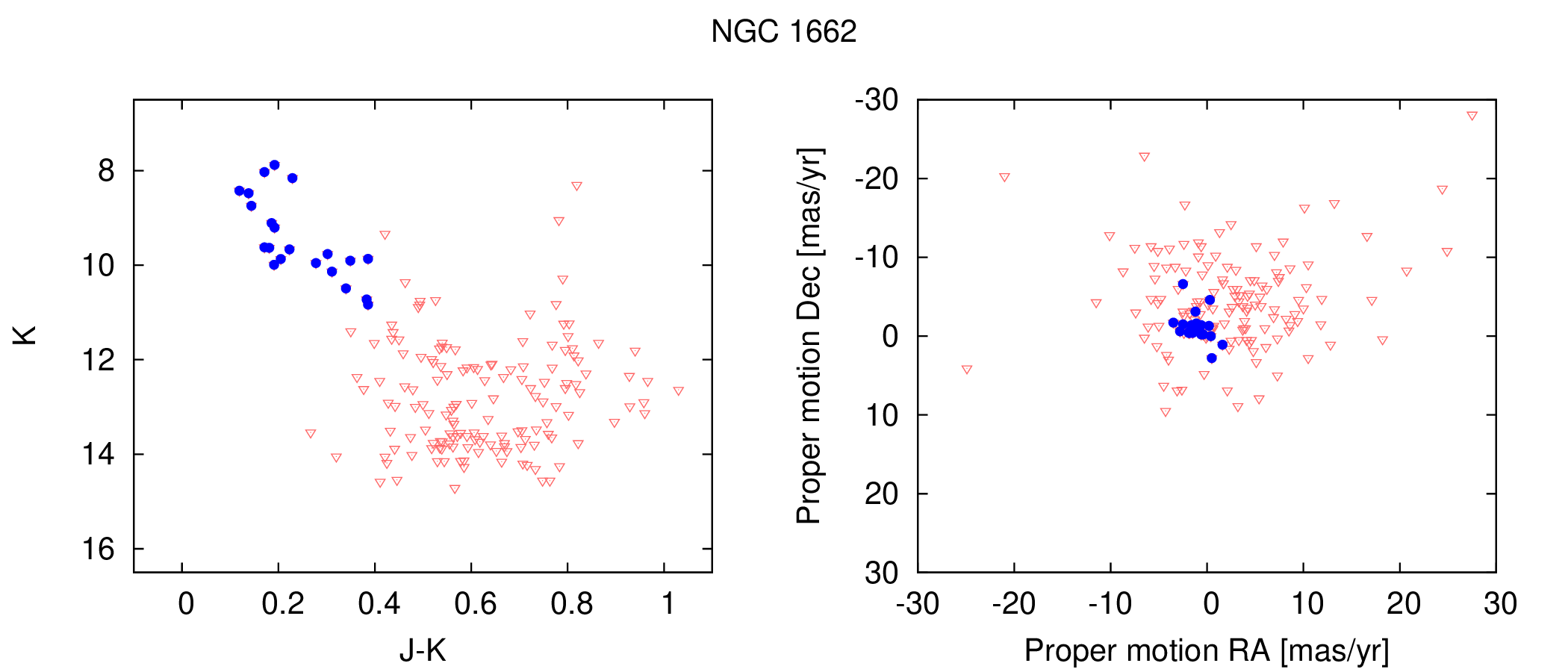}
\includegraphics[width=14cm]{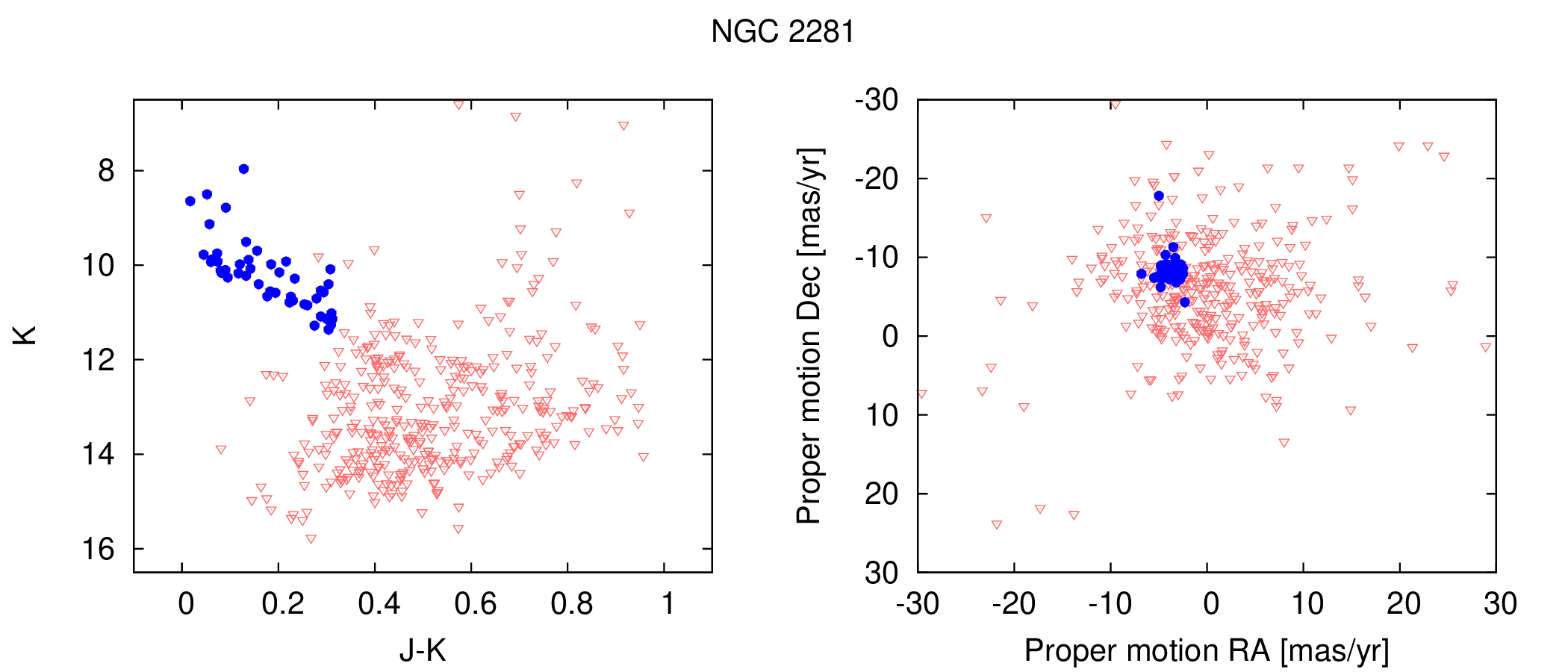}
\end{figure}
\begin{figure}
\centering
\includegraphics[width=14cm]{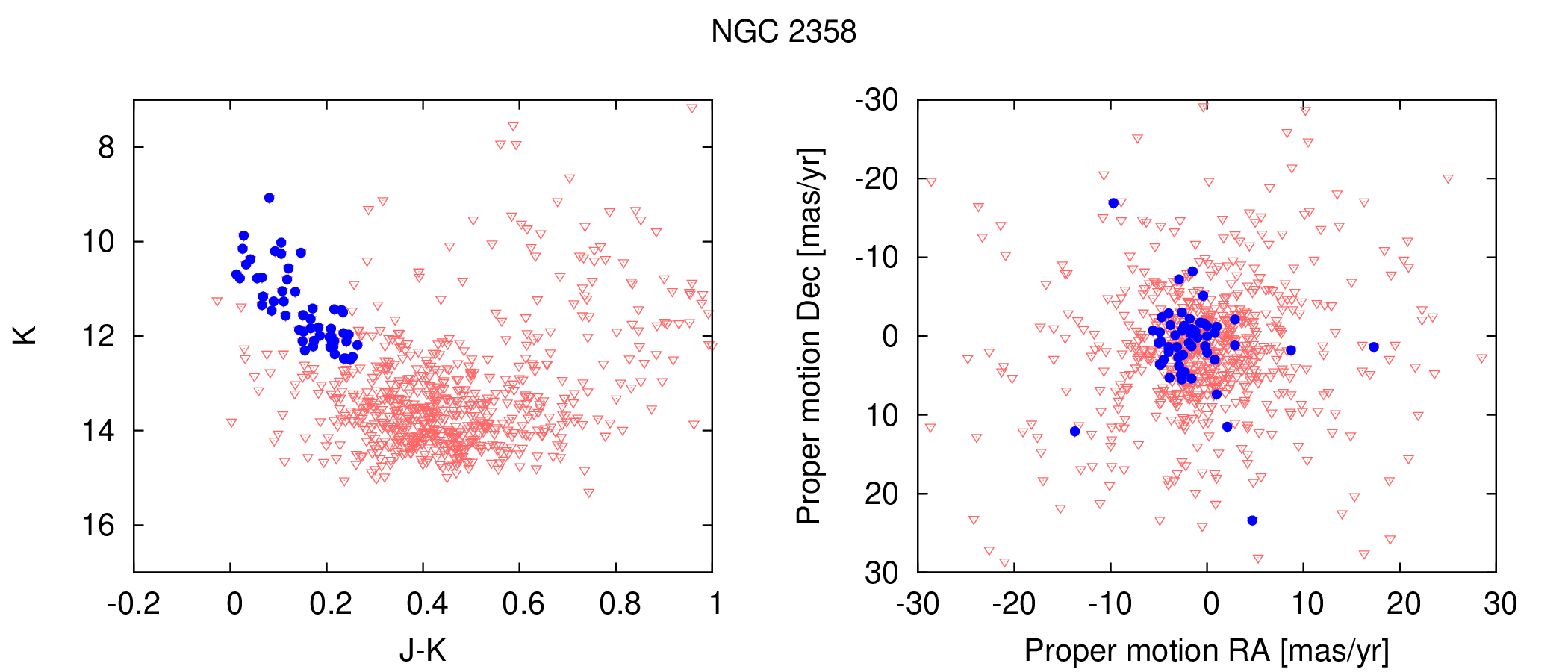}
\includegraphics[width=14cm]{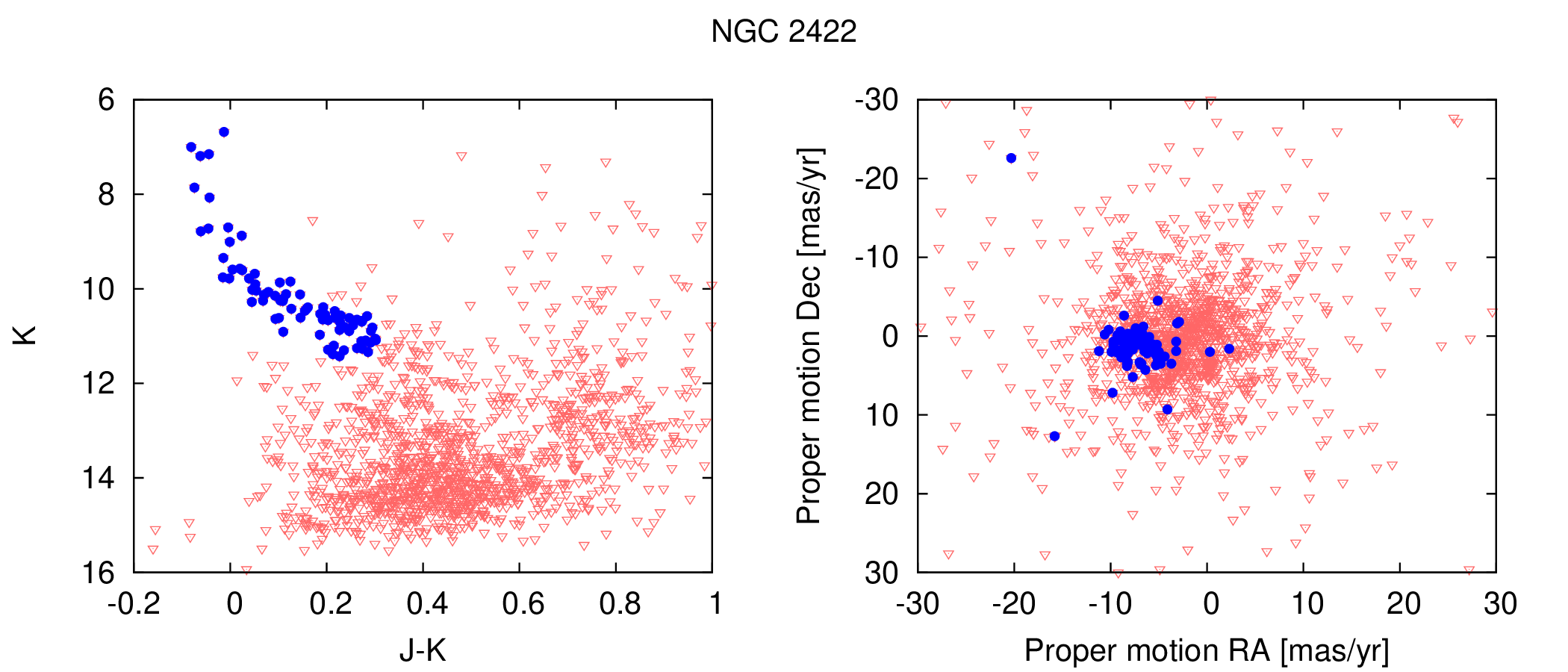}
\includegraphics[width=14cm]{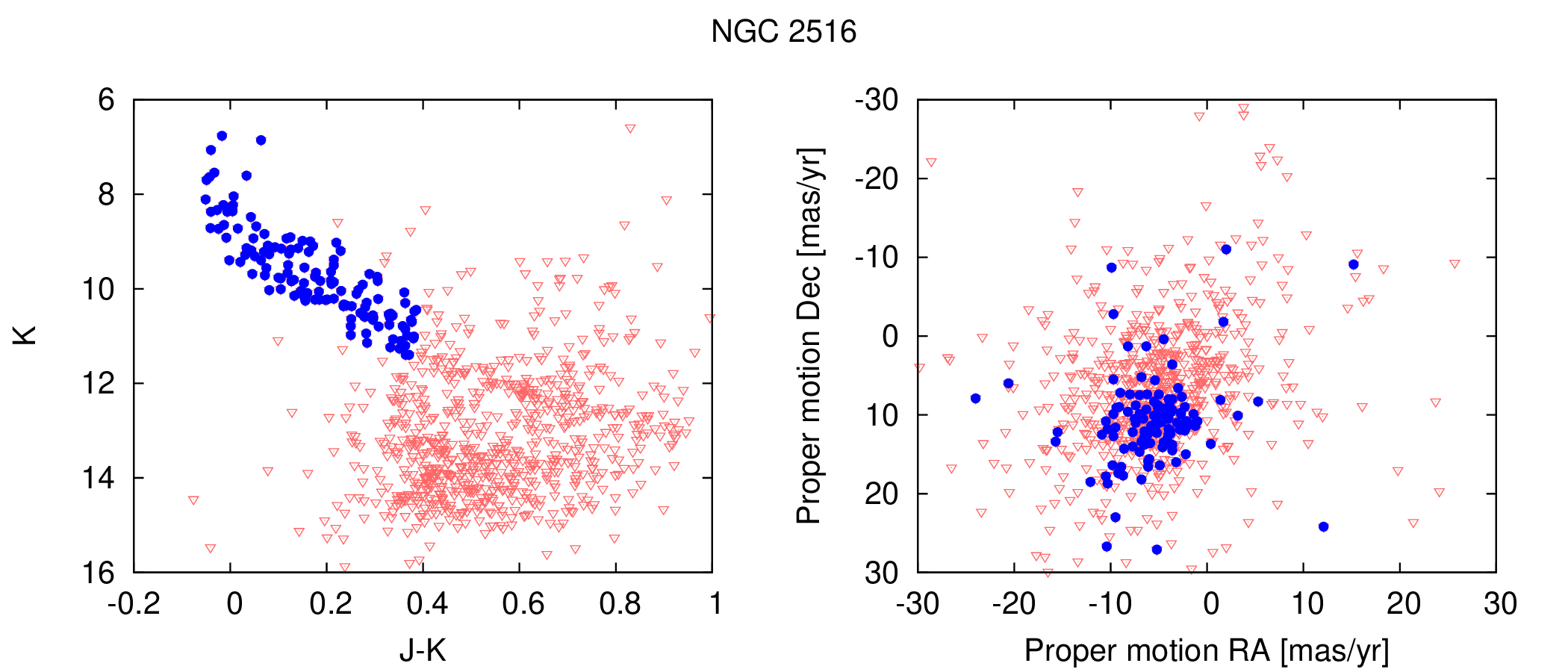}
\includegraphics[width=14cm]{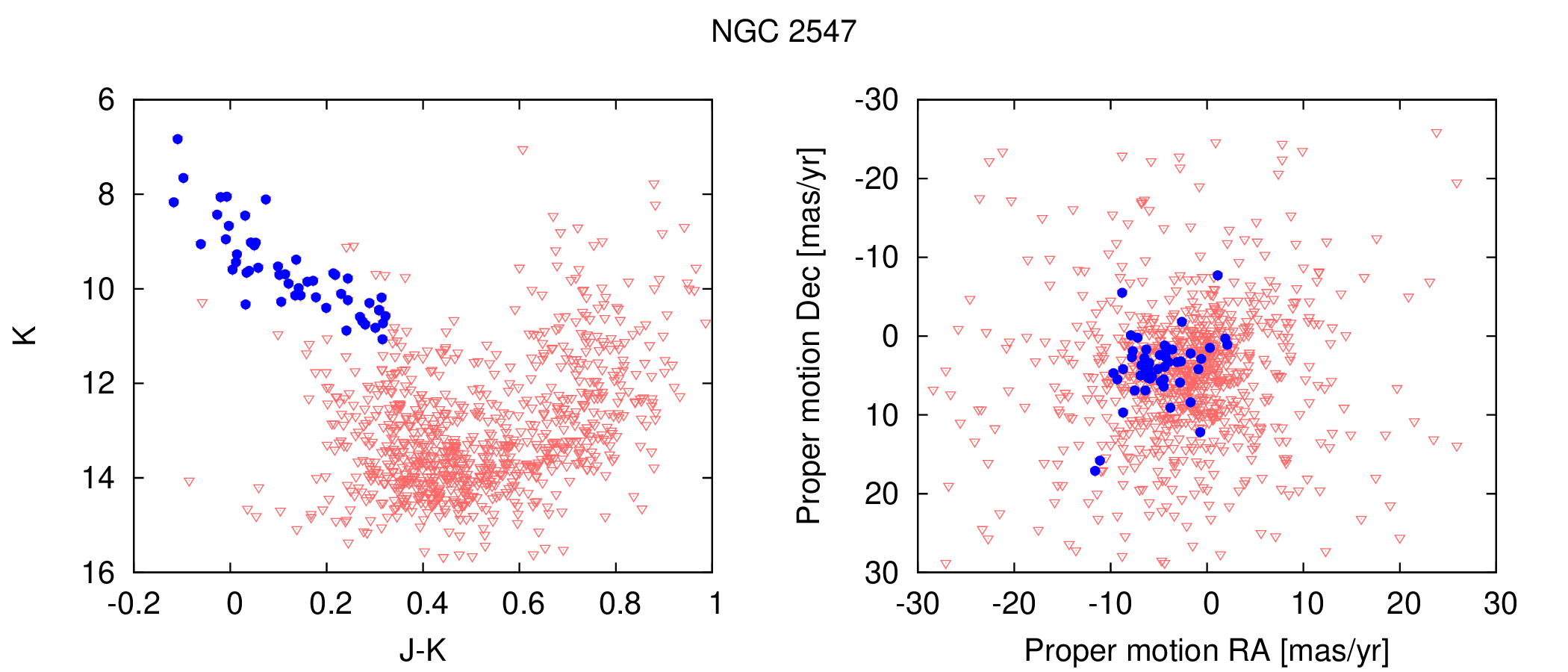}
\end{figure}
\begin{figure}
\centering
\includegraphics[width=14cm]{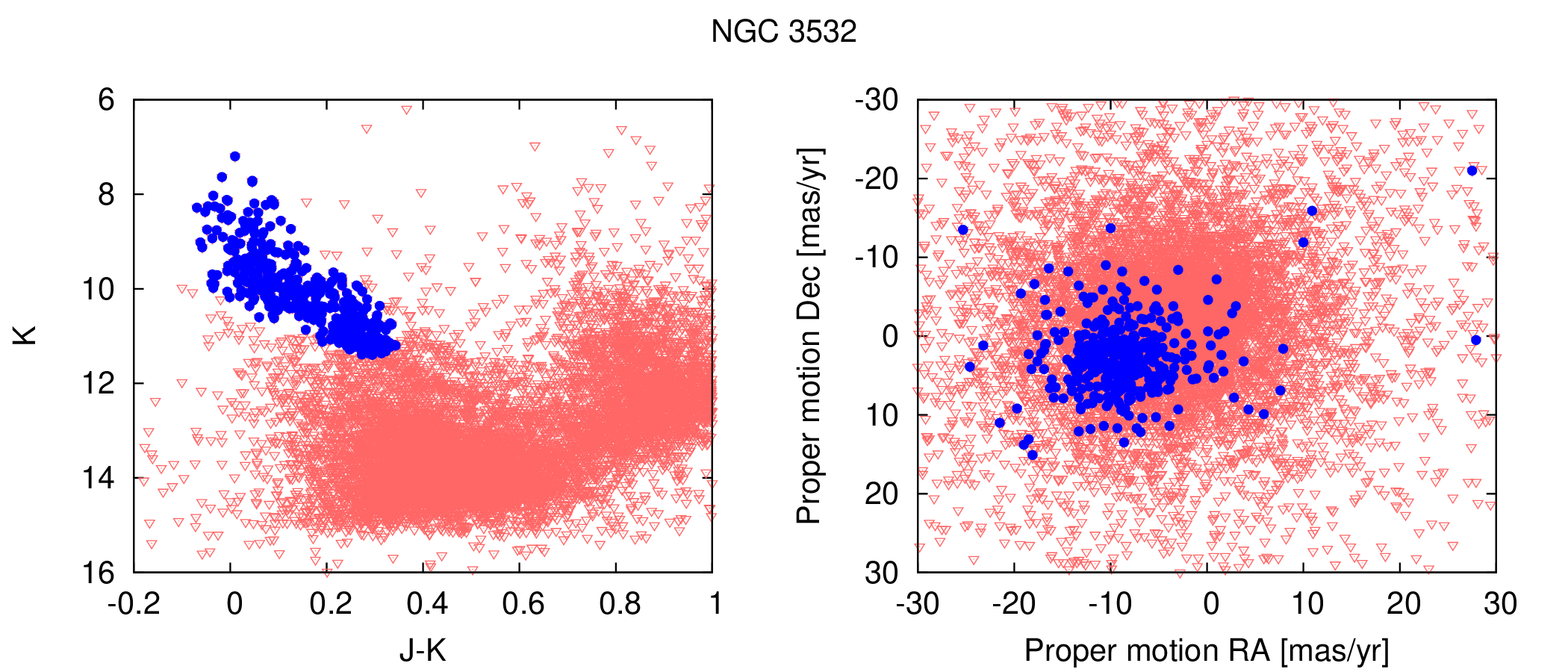}
\includegraphics[width=14cm]{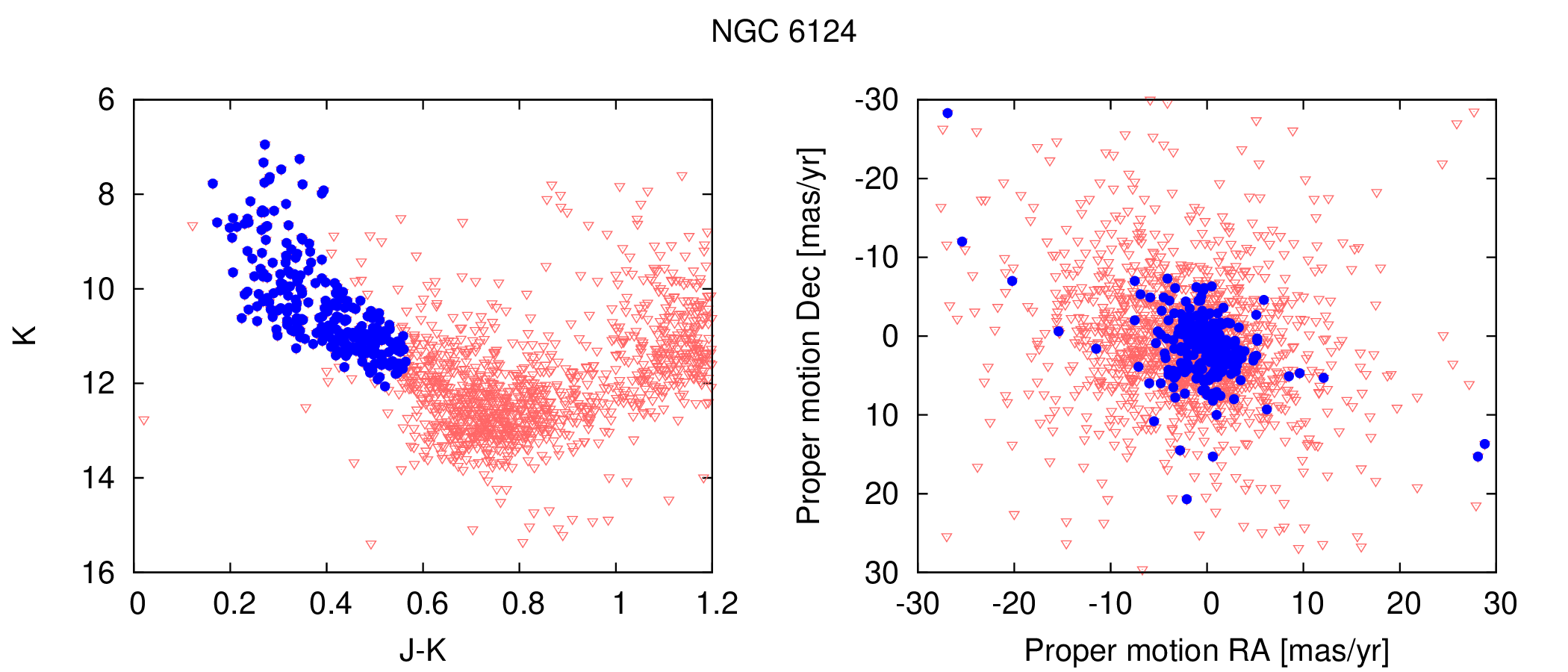}
\includegraphics[width=14cm]{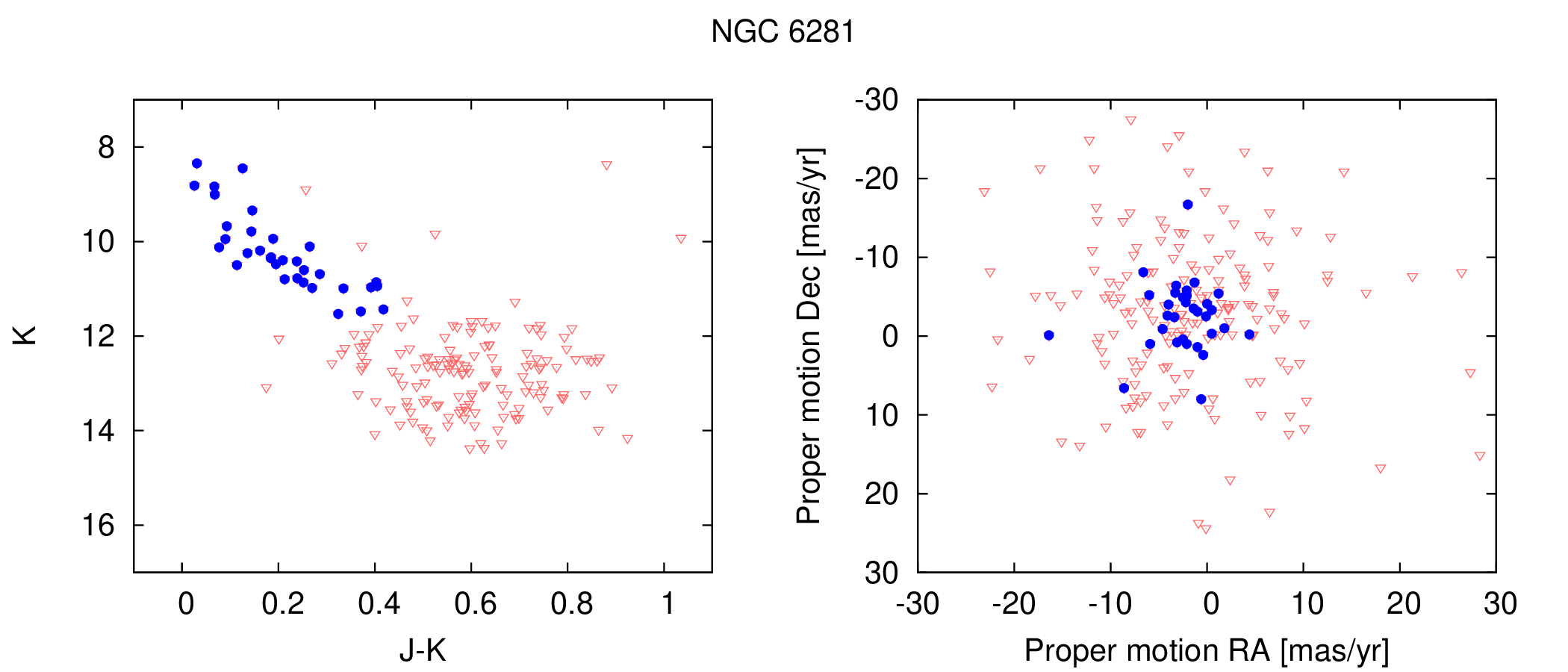}
\includegraphics[width=14cm]{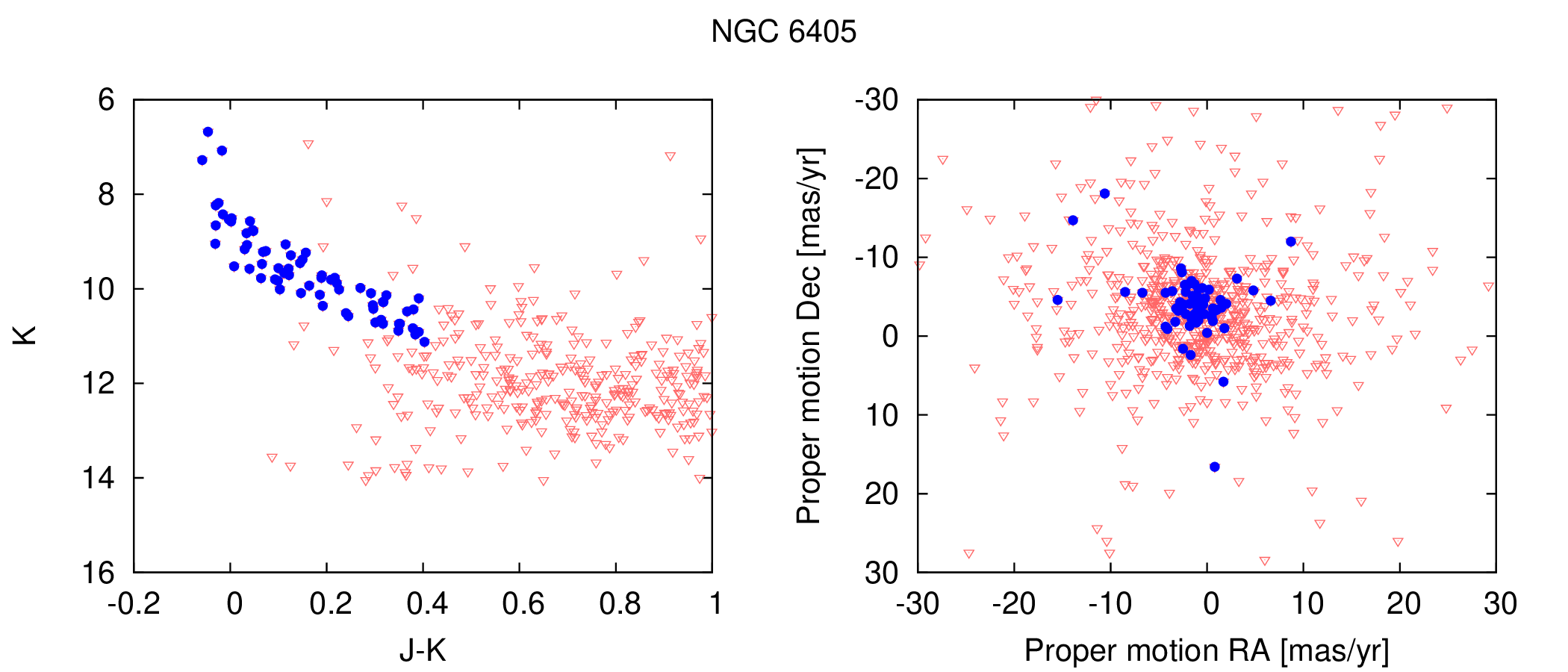}
\end{figure}

Outlying points in the $N_{1}$ subselections were removed using median absolute deviation (MAD), defined as:

\begin{equation} \label{mad}
\begin{split}
MAD(x) = median_{i}(|x_{i}-median_{j}(x_{j})|) \\
MAD(\mu)=\sqrt{(MAD(\mu_{\alpha}cos\delta))^{2}+(MAD(\mu_{\delta}))^{2}}
\end{split}
\end{equation}

The value of $MAD(\mu)$ was calculated for each cluster. Sources with proper motion differing by more than $4MAD(\mu)$ from the median proper motion were considered outliers and excluded from the sample, thus producing even narrower subselections consisting of $N_{2}$ stars. The proper motions of the clusters were finally calculated by averaging the data in the $N_{2}$ subselections.

\section{Results}

\begin{figure}
\centering
\includegraphics[width=9.3cm]{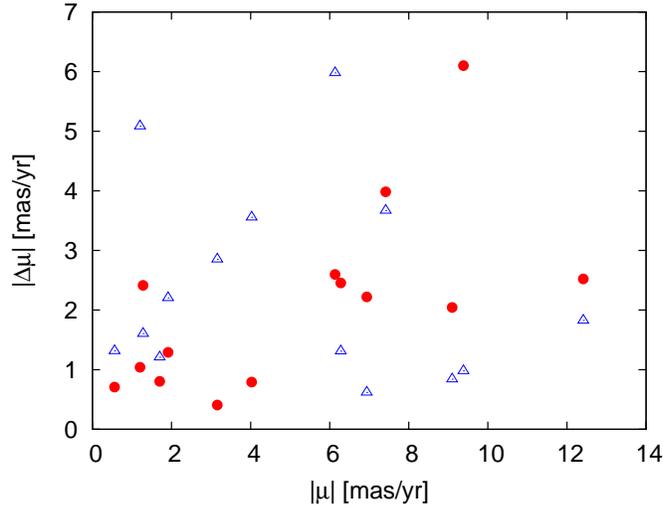}
\caption{Comparison of the calculated proper motions with the values by \citet{2014A&A...564A..79D} \textit{(red circles)} and \citet{2003ARep...47....6L} \textit{(blue triangles)}. The x-axis represents absolute proper motion values, calculated in this work, while the y-axis represents absolute values of vector differences to the previous estimates. The \citet{2014A&A...564A..79D} data point for NGC 7092 lies outside the plot, at $(|\mu|,|\Delta\mu|)=(18.18,17.08)$.}
\label{comparison}
\end{figure}

Our results are presented in Table\,\ref{results}. The standard deviations of the proper motions in the $N_{2}$ subselections are in the range of $0.8$\,mas/yr$-4$\,mas/yr, which is comparable to the errors given by \citet{2014A&A...564A..79D}. However, the results differ significantly from theirs ($|\Delta\mu|>2$\,mas/yr for 9 of the 15 clusters). Very large deviations are observed for NGC 7092, NGC 3532 and NGC 2422. Higher deviations from \citet{2014A&A...564A..79D} are generally observed at higher absolute proper motion values (Fig.\,\ref{comparison}). 

We suggest that \citet{2014A&A...564A..79D} may have used a large number of background stars, which could have contaminated their selections. We attempted to estimate the percentage of those background stars. For each cluster we examined 4 nearby fields, centered 40$'$ away (60$'$ away in the case of the larger NGC 3532), and with radius $r_{s}$, equal to the search radius for the cluster (Table\,\ref{results}). The median number $N_{F}$ of UCAC4 sources in these 4 fields was then calculated. The portion of field stars should be roughly $f=N_{F}/N_{0}$. For all clusters $f>67\%$. The portion of field stars among those used by \citet{2014A&A...564A..79D} would be approximately $f_{D}=1-(1-f)N_{0}/N_{D}$. The minimum and median values of $f_{D}$ are 57\% and 75\% respectively. Although this is just a rough estimate, it shows that a considerable portion of stars used by \citet{2014A&A...564A..79D} are not physical members of the respective clusters.

\citet{2003ARep...47....6L} have also applied photometric criteria for their selections. Our agreement with the latter is slightly better in general (median $|\Delta\mu|$ of 1.6\,mas/yr ) and much better in the case of NGC 7092 ($|\Delta\mu|=1.52$\,mas/yr and 17.08\,mas/yr when comparing the data in Table\,\ref{results} to \citet{2003ARep...47....6L} and \citet{2014A&A...564A..79D} respectively). The proper motion diagram for NGC 7092 (Fig.\,\ref{CMDs}) contains a considerable number of outlying points. The reason is that NGC 7092 is a very close cluster, located near the galactic plane (Table\,\ref{table_objects}). Most of the outliers are not in the $N_{2}$ subselection and do not affect the result as they lie farther than $4MAD(\mu)$ from the median value.

\begin{table}
\begin{center}
\caption[.]{Proper motions calculated for 15 open clusters. The last column contains the number of stars used by \citet{2014A&A...564A..79D}. }\label{results}
\small
\begin{tabular}{l@{ }c@{ }r@{ }r@{ }c@{ }r@{ }c@{ }c@{ }c@{ }c@{ }r@{ }}
\hline\hline
cluster \, & $r_{s}$ \,& $N_{0}$ \,& $N_{1}$ \,&  $MAD(\mu)$ \,& $N_{2}$ \,&  $\mu_{\alpha}cos\delta$ \,& $\sigma_{\alpha}$ \,& $\mu_{\delta}$ \,& $\sigma_{\delta}$ \,& $N_{D}$ \\ 
name \, & [arcmin] \,&   \,&   \,&  [mas/yr] \,&   \,&  [mas/yr] \,& [mas/yr] \,& [mas/yr] \,& [mas/yr] \,&   \\   
\hline
\object{NGC 1039} \,&   18.5 \,&  1022  \,&     86 \,&   0.92 \,&     72  \,& 	-0.56 \,&  1.03 \,&    -6.26 \,&  0.82 \,&  783 \\
\object{NGC 1647} \,&   21.0 \,&   848  \,&     87 \,&   1.14 \,&     78  \,& 	-1.13 \,&  1.35 \,&    -1.27 \,&  1.24 \,&  656 \\
\object{NGC 1662} \,&   11.0 \,&   173  \,&     21 \,&   0.99 \,&     19  \,& 	-1.10 \,&  1.24 \,&    -0.66 \,&  1.21 \,&  151 \\
\object{NGC 2281} \,&   13.5 \,&   439  \,&     46 \,&   0.92 \,&     43  \,& 	-3.92 \,&  0.91 \,&    -8.21 \,&  0.92 \,&  330 \\
\object{NGC 2358} \,&   11.0 \,&   750  \,&     55 \,&   2.83 \,&     49  \,& 	-1.85 \,&  2.56 \,&     0.49 \,&  3.10 \,&  618 \\
\object{NGC 2422} \,&   13.5 \,&  1487  \,&     78 \,&   1.64 \,&     73  \,& 	-7.29 \,&  1.87 \,&     1.38 \,&  1.79 \,& 1293 \\
\object{NGC 2516} \,&   16.0 \,&   941  \,&    134 \,&   2.84 \,&     117 \,& 	-5.48 \,&  3.13 \,&    11.14 \,&  3.36 \,&  737 \\
\object{NGC 2547} \,&   13.5 \,&   960  \,&     51 \,&   2.55 \,&     48  \,& 	-4.88 \,&  2.80 \,&     3.71 \,&  2.96 \,&  644 \\
\object{NGC 3532} \,&   26.0 \,& 11974  \,&    409 \,&   3.40 \,&     386 \,& 	-8.90 \,&  3.91 \,&     2.97 \,&  3.80 \,& 8705 \\
\object{NGC 6124} \,&   20.5 \,&  1838  \,&    263 \,&   2.72 \,&     243 \,& 	-0.18 \,&  2.49 \,&     1.19 \,&  3.16 \,& 1633 \\
\object{NGC 6281} \,&    5.0 \,&   280  \,&     33 \,&   2.83 \,&     30  \,& 	-1.92 \,&  2.40 \,&    -2.51 \,&  3.40 \,&  207 \\
\object{NGC 6405} \,&   11.0 \,&   930  \,&     67 \,&   2.05 \,&     61  \,& 	-1.11 \,&  2.33 \,&    -3.87 \,&  2.12 \,&  737 \\
\object{NGC 6494} \,&   15.5 \,&  1640  \,&    185 \,&   2.36 \,&     162 \,&     0.49 \,&  2.80 \,&    -0.27 \,&  2.37 \,& 1342 \\
\object{NGC 7092} \,&   15.5 \,&  2019  \,&     34 \,&   2.77 \,&     25  \,& 	-8.20 \,&  1.18 \,&   -18.14 \,&  3.97 \,& 1464 \\
\object{IC 4725}  \,&   15.5 \,&  5812  \,&    124 \,&   2.84 \,&     111 \,& 	-3.46 \,&  2.55 \,&    -6.01 \,&  3.76 \,& 4458 \\
\hline
\end{tabular} 
\end{center}  
\end{table}  

\section{Summary}
\label{sect:summary}

Proper motions are important parameters of open clusters, which help us improve our understanding of galactic dynamics. We built NIR color-magnitude diagrams of 15 open clusters and we used them to select stars that are very probable members. After excluding the ones with an uncommon proper motion, we used those subselections to calculate the proper motions of the clusters. Our results suggest that \citet{2014A&A...564A..79D} may have used selections, contaminated by background stars. Our work shows the advantage of utilizing CMDs for the calculation of open cluster proper motions.

\normalem

\begin{acknowledgements}
This work was supported by grant No. BG051 PO001-3.3.06-0057 of the European Social Fund. It was carried out partly during the 2014 Beli Brezi Summer School of Astronomy and Astrophysics, organized by the Kardzhali Astronomical Observatory and the University of Sofia, Bulgaria.

\end{acknowledgements}
  
\bibliographystyle{raa}
\bibliography{clusters_pm}

\end{document}